\documentclass[aps,nofootinbib,prd,eqsecnum,twocolumn,showpacs,showkeys,preprintnumbers]{revtex4-1}
\usepackage{graphicx}
\usepackage{lineno}
\usepackage{graphicx}
\usepackage{amsmath}
\usepackage{amsfonts}
\usepackage{amssymb}
\usepackage{color}
\usepackage{bm}
\usepackage{float}
\usepackage{mathrsfs}
\usepackage{epstopdf}
\usepackage{url}
\usepackage{placeins}
\usepackage{footnote}
\usepackage{textcomp}
\usepackage[normalem]{ulem}
\usepackage{xcolor}
\usepackage[unicode=true, pdfusetitle,
 bookmarks=true,bookmarksnumbered=false,bookmarksopen=false,
 breaklinks=false,pdfborder={0 0 1},backref=false,colorlinks=false]{hyperref}
\usepackage{multirow}
\usepackage{pifont}
\usepackage{times}
\usepackage[english]{babel}

\usepackage{adjustbox}
\usepackage{amsmath}
\usepackage{booktabs}
\usepackage{caption}

\usepackage{float}

\usepackage{tabularx}
\makeatletter

\newcommand{\stkout}[1]{\ifmmode\text{\sout{\ensuremath{#1}}}\else\sout{#1}\fi}

\newcolumntype{L}[1]{>{\hsize=#1\hsize\raggedright\arraybackslash}X}%
\newcolumntype{R}[1]{>{\hsize=#1\hsize\raggedleft\arraybackslash}X}%
\newcolumntype{C}[1]{>{\hsize=#1\hsize\centering\arraybackslash}X}%

\newcommand*\patchAmsMathEnvironmentForLineno[1]{%
 \expandafter\let\csname old#1\expandafter\endcsname\csname #1\endcsname
 \expandafter\let\csname oldend#1\expandafter\endcsname\csname end#1\endcsname
 \renewenvironment{#1}%
   {\linenomath\csname old#1\endcsname}%
   {\csname oldend#1\endcsname\endlinenomath}}%
\newcommand*\patchBothAmsMathEnvironmentsForLineno[1]{%
 \patchAmsMathEnvironmentForLineno{#1}%
 \patchAmsMathEnvironmentForLineno{#1*}}%
\AtBeginDocument{%
\patchBothAmsMathEnvironmentsForLineno{align}%
\patchBothAmsMathEnvironmentsForLineno{flalign}%
\patchBothAmsMathEnvironmentsForLineno{alignat}%
\patchBothAmsMathEnvironmentsForLineno{gather}%
\patchBothAmsMathEnvironmentsForLineno{multline}%
}

\begin{document}
\title{
Testing $f(Q)$ Gravity with Logarithmic Equation of State Using Latest Cosmological Data}
\author{Chaymae Karam $^{1}$}
\email{chaymae$_$karam2@um5.ac.ma}
\author{Dalale Mhamdi$^{2,3}$}
\email{dalale.mhamdi@ump.ac.ma}
\author{Taoufik Ouali$^{2,3}$}
\email{t.ouali@ump.ac.ma}
\author{Rachid Ahl Laamara$^{1}$}
\email{r.ahllaamara@um5r.ac.ma}
\author{Mohamed Bennai$^{1,4}$}
\email{mohamed.bennai@univh2c.ma}
\date{\today }

\affiliation{$^{1}$Laboratory of High Energy Physics, Modeling and Simulations, Faculty of Science, University Mohammed V in Rabat, Rabat, Morocco.\\$^{2}$Laboratory of Physics of Matter and Radiation, University of Mohammed first, BP 717, Oujda, Morocco.\\$^{3}$Astrophysical and Cosmological Center, Faculty of Sciences, University of Mohammed first, BP 717, Oujda, Morocco.\\$^{4}$Quantum Physics and Spintronic Team, LPMC, Faculty of Sciences Ben M’sick, Hassan II University of Casablanca, Morocco.
\\}


\begin{abstract}
In this paper, we investigate the late-time accelerated expansion of the Universe in power law $f(Q)$ gravity, where a logarithmic dark energy parametrization is considered: $\omega_{de}(z)=\omega_0+\omega_1\ln(1+z)$. This description gives a smooth deviation from a constant equation of state within the complete range of redshifts. We obtain an analytical expression for the Hubble parameter and we use a Markov Chain Monte Carlo to compare the model with the most recent observational data obtained from Pantheon$^+$ Type Ia supernovae, Baryon Acoustic Oscillation measurements from the second data release (DR2) of the Dark Energy Spectroscopic Instrument, and Cosmic Chronometers. We then perform a statistical comparison between our model and the standard $\Lambda$CDM model using the Akaike Information Criterion and the Bayesian Information Criterion. From our results, we conclude that the use of logarithmic parametrization in the $f(Q)$ gravity model $f(Q)=Q+6\gamma H_0^2\left(\frac{Q}{Q_0}\right)^n$ is a valid and more flexible alternative to standard dark energy models, as it provides a richer phenomenology at low redshifts.
\end{abstract}

\maketitle

\section{Introduction}

Over the past decades, the observational discovery of the accelerated expansion of the Universe \cite{ref1,ref2} has profoundly challenged our understanding of gravitation and cosmic dynamics. Explaining this phenomenon remains one of the major challenges in modern cosmology. Two main theoretical approaches have been explored to account for the late-time accelerated expansion of the Universe: either introducing a new energy component, commonly referred to as dark energy (DE) or modifying Einstein’s general relativity. In the standard framework, this dark energy is typically modeled as a cosmological constant, $\Lambda$, characterized by a negative pressure that produces a repulsive gravitational effect on cosmological scales. Together with dark matter, which is responsible for the formation of large-scale structures, this constant forms the cornerstone of the $\Lambda$CDM model, which remains the most widely supported scenario by current observations.
However, despite its remarkable success, the $\Lambda CDM$ model is not free from theoretical difficulties. It faces two major problems related to the nature and origin of the cosmological constant \cite{ref3,ref4,ref5,ref6,ref7,ref8,ref9,ref10}, as well as growing tensions between different measurements of the current expansion rate of the Universe. These limitations have motivated the exploration of alternative avenues, including dynamic forms of dark energy or modified theories of gravity.
The equation of state parameter $\omega(z)$, defined as the ratio of a fluid’s pressure to its energy density, is widely used to characterize the dynamics of dark energy. The simplest way to explain cosmological observations is to consider dark energy as a cosmological constant, corresponding to a redshift independent equation of state parameter, $\omega =-1$. Some introduce non-standard forms of dark energy within the framework of general relativity, including quintessence \cite{ref11,ref12}, phantom models \cite{ref7,ref13,ref14,ref15,ref16,ref17,ref18}, k-essence \cite{ref19,ref20}, and holographic dark energy \cite{ref21,ref22,ref23,ref24,ref25,ref26,ref27,ref28}. Others focus on modifying the gravitational sector of Einstein’s equations, giving rise to modified gravity theories \cite{ref29,ref30,ref31,ref32,ref33,ref34,ref35,ref36,ref37,ref38,ref39}
, anisotropic cosmological models \cite{ref40,ref41,ref42}, or scenarios involving interactions between dark matter and dark energy \cite{ref43,ref44,ref45}. These extensions seek to account for the cosmic acceleration and other unresolved phenomena by providing mechanisms alternative to conventional dark energy, while remaining compatible with current observations \cite{ref46,ref47,ref48,ref49,ref50,ref51,ref52,ref53}.
In classical general relativity, gravity is understood as a manifestation of the curvature of spacetime, a view rooted in the equivalence principle. Within this framework, the dynamics of gravitation are fully determined by the distribution of matter energy density encoded in the energy-momentum tensor, the Ricci tensor and the scalar curvature, which play a central role in describing the underlying Riemannian geometry, where both torsion and non-metricity vanish. Although this formulation has been remarkably successful in explaining gravitational phenomena especially within the Solar System it faces significant challenges at cosmological and galactic scales. The late-time accelerated expansion of the Universe and the dynamics typically attributed to dark matter suggest that Einstein’s theory, in its standard form, may be incomplete when applied beyond local gravitational systems.
These limitations have motivated the development of several extensions of general relativity, either by enriching the Einstein–Hilbert Lagrangian or by modifying its geometric foundations. Such approaches have led to numerous important extensions, including $f(R)$ gravity \cite{ref54}, $f(G)$ gravity \cite{ref55}, $f(P)$ gravity \cite{ref56}, and Horndeski scalar-tensor theories \cite{ref57}, etc. From a broader geometric perspective, however, curvature is not the only quantity capable of encoding gravitational effects: torsion and non-metricity also constitute fundamental attributes of a general metric affine geometry. Incorporating these structures naturally gives rise to the teleparallel $f(T)$ theories \cite{refA}, based on torsion, and the symmetric teleparallel equivalent of general relativity (STEGR) described by $f(Q)$ theories \cite{refB}, where the non-metricity scalar $Q$ governs the gravitational interaction. These formalisms offer promising avenues for revisiting the gravitational sector and exploring alternatives to dark energy within a more general geometric framework \cite{ref95}. Indeed, $f(Q)$ gravity has recently emerged as a promising alternative within the symmetric teleparallel framework. A key advantage of this theory is that it leads to second order field equations, avoiding the higher order instabilities often encountered in curvature based extensions. In addition, its geometric formulation provides a natural mechanism to account for the late-time acceleration of the Universe without invoking an explicit dark energy component, making it particularly suitable for cosmological applications. Besides, these kind of gravities distinguish from general relativity at perturbative levels particularilly at the growth of  structures.

In this work, we examine the late-time acceleration of the Universe within the framework of symmetric teleparallel gravity, adopting a power law form of $f(Q)$ originally proposed in Refs \cite{ref58,ref59}. In this geometrical setting, gravitation is described through the non-metricity scalar, $Q$, in a flat and curvature-free spacetime. We consider the specific model $f(Q)=Q + 6\gamma H_0^{2}\!\left(\frac{Q}{Q_0}\right)^{n}$ \cite{ref92} with $\qquad Q_0 = 6H_0^{2}$, from which we construct an effective dark-energy component emerging naturally from the modified Friedmann equations. To allow deviations from the standard cosmological constant scenario, we employ a logarithmic parameterization of the dark energy equation of state, $\omega_{\rm de}(z) = \omega_{0} + \omega_{1}\ln(1+z)$, which provides a smooth departure from $\omega=-1$ across all redshifts \cite{ref63}.

We derive an analytical expression for the Hubble expansion rate and confront the model with current observations by combining Pantheon$^+$ Type Ia supernovae (SNIa), Baryon Acoustic Oscillation (BAO) measurements from the second data release (DR2) of the Dark Energy Spectroscopic Instrument (DESI), and Cosmic Chronometers (CC). A Markov Chain Monte Carlo analysis is performed to constrain the parameter space $(h,\,\Omega_{m0},\,\omega_{0},\,\omega_{1},\,n)$, allowing us to assess the observational viability of the theory. We subsequently assess our model against the standard $\Lambda$CDM framework using the Akaike Information Criterion (AIC)~\cite{ref70} and the Bayesian Information Criterion (BIC)~\cite{ref71}. We further analyze several cosmological diagnostics including the deceleration, jerk, snap, dark matter and dark energy densities, $Om(z)$ diagnostic, and statefinder parameters to probe the dynamical behavior of the model. Our results show that the logarithmic parametrization within this power law $f(Q)$ framework provides a flexible alternative to standard dark energy scenarios, offering a richer low redshift phenomenology and the potential to mitigate existing cosmological tensions.

The manuscript is structured as follows. In Section~\ref{sec:1}, we present the basic formalism of $f(Q)$ gravity and the assumptions underlying the construction of the specific cosmological model. In Section~\ref{sec:3}, we present the dynamics of power law $f(Q)$ gravity with a logarithmic equation of state for dark energy. In Section~\ref{sec:4}, we describe the observational data and statistical methodology. Section~\ref{sec:5} is devoted to the numerical results and discussion. In Section~\ref{sec:6}, we examine the evolution of key cosmological quantities such as the deceleration, jerk, and snap parameters, the dark matter and dark energy densities, as well as the $Om(z)$ and statefinder diagnostics. Finally, in Section~\ref{sec:7}, we present a summary of the main results and our concluding remarks.

\section{Overview of $f(Q)$ Gravity}
\label{sec:1}

The $f(Q)$ gravity represents an extension of the symmetric teleparallel formulation of general relativity, defined on a spacetime with vanishing curvature and torsion \cite{ref60,ref61}. In differential geometry, particularly within Weyl-Cartan geometry the symmetric metric tensor $g_{\mu\nu}$ is used to measure the length of vectors, while an asymmetric connection $\Sigma^\gamma{ }_{\mu \nu}$ determines covariant derivatives and parallel transport. In this framework, the general affine connection can be naturally decomposed into three components: the Christoffel symbol $\Gamma^\gamma{ }_{\mu \nu}$, corresponding to the Levi-Civita part, the contortion tensor $C^\gamma{ }_{\mu \nu}$ related to torsion, and the disformation tensor $L^\gamma{ }_{\mu \nu}$ describing non-metricity \cite{ref62}

\begin{equation}
\Sigma^\gamma{ }_{\mu \nu}=\Gamma^\gamma{ }_{\mu \nu}+C^\gamma{ }_{\mu \nu}+L^\gamma{ }_{\mu \nu}.
 \label{1}
\end{equation}
 
The Levi-Civita connection of the metric, $g_{\mu \nu}$, is defined in its usual form

\begin{equation}
\Gamma^\gamma{ }_{\mu \nu} \equiv \frac{1}{2} g^{\gamma \sigma}\left(\frac{\partial g_{\sigma \nu}}{\partial x^\mu}+\frac{\partial g_{\sigma \mu}}{\partial x^\nu}-\frac{\partial g_{\mu \nu}}{\partial x^\sigma}\right).  
\end{equation}
 
The contorsion tensor $C^\gamma{ }_{\mu \nu}$ can be expressed as

\begin{equation}
C^\gamma{ }_{\mu \nu} \equiv \frac{1}{2} T^\gamma{ }_{\mu \nu}+T_{(\mu}{ }^\gamma{ }_{\nu)}.   \label{2}
\end{equation}

In Eq.~(\ref{2}), $T^\gamma{ }_{\mu \nu} \equiv 2 \Sigma^\gamma{ }_{[\mu \nu]}$ represents the torsion tensor. Lastly, the disformation tensor $L^\gamma{ }_{\mu \nu}$ is obtained from the non-metricity tensor $Q_{\gamma \mu \nu}$ as follows

\begin{equation}
L^\gamma{ }_{\mu \nu} \equiv - \frac{1}{2} g^{\gamma \sigma}\left(Q_{\mu \sigma \nu}+Q_{\nu \sigma \mu}-Q_{\sigma \mu \nu}\right) .   
\end{equation}
 
In the equation provided, the non-metricity tensor $Q_{\gamma \mu \nu}$ is defined as the covariant derivative of the metric tensor with respect to the Weyl-Cartan connection $\Sigma^\gamma{ }_{\mu \nu}$, represented as $Q_{\gamma \mu \nu}=\nabla_\gamma g_{\mu \nu}$.

The connection under consideration is assumed to be devoid of torsion and curvature in the current background. It represents a simple coordinate transformation from the trivial connection mentioned in \cite{ref88}. Therefore, for a connection that is both flat and free of torsion, it can be parameterized as follows
\begin{equation}
\Sigma^\gamma{ }_{\mu \beta}=\frac{\partial x^\gamma}{\partial \xi^\rho} \partial_\mu \partial_\beta \xi^\rho . 
\end{equation}
where $\xi^\gamma=\xi^\gamma\left(x^\mu\right)$ is an invertible relation. It is always possible to select a coordinate system where the connection becomes zero. This condition is referred to as the coincident gauge and has been employed in many studies of STEGR \cite{ref62}.
The Symmetric teleparallel gravity provides a geometric description of gravity that is equivalent to General Relativity (STEGR) when using coincident gauge coordinates, i.e., $\Sigma^\gamma{ }_{\mu \nu}=0$ and $C^\gamma{ }_{\mu \nu}=0$. Therefore, from \eqref{1}, we can infer that \cite{ref93}

\begin{equation}
\Gamma^\gamma{ }_{\mu \nu}=-L^\gamma{ }_{\mu \nu} . 
\end{equation} 
 
In $f(Q)$ gravity, the gravitational action is modified by replacing the Ricci scalar R in the Einstein-Hilbert action with a function $f(Q)$, where $Q$ is the non-metricity scalar. This scalar encapsulates the geometric property of non-metricity, allowing for a new interpretation of gravitational interactions that can potentially address cosmological phenomena such as dark energy and the accelerated expansion of the Universe. The action for $f(Q)$ gravity is given by

\begin{equation}
 S=\int\left[\frac{1}{2} f(Q)+\mathcal{L}_m\right] d^4 x \sqrt{-g}            \label{7}
 \end{equation}
where $\mathcal{L}_m$ is the matter Lagrangian density, and $g$ is the determinant of the metric tensor $g_{\mu \nu}$. We have set $8 \pi G=1$.
To derive the field equations, we vary this action with respect to the metric $g_{\mu \nu}$. 
The non-metricity tensor $Q_{\gamma\mu\nu}$ is defined as
\begin{equation}
 \begin{gathered}
Q_{\gamma \mu \nu}=\nabla_\gamma g_{\mu \nu}, \\
Q_\gamma=Q_\gamma{ }^\mu{ }_\mu, \quad \tilde{Q}_\gamma=Q^\mu{ }_{\gamma \mu},
\end{gathered}   
\end{equation} 
here, $\nabla$ denotes the covariant derivative. The superpotential tensor, also known as the non-metricity conjugate, can be expressed as

\begin{equation}
4 P^\gamma{ }_{\mu \nu}=-Q^\gamma{ }_{\mu \nu}+2 Q_{(\mu}{ }^\gamma{ }_{\nu)}+Q^\gamma g_{\mu \nu}-\widetilde{Q}^\gamma g_{\mu \nu}-\delta_{(\mu}^\gamma Q_{\nu)},
\end{equation}  
where the trace of the non-metricity tensor is given by

\begin{equation}
Q=-Q_{\gamma \mu \nu} P^{\gamma \mu \nu}.
\end{equation}

The variation of the action, Eq.~(\ref{7}), with respect to the metric leads to the field equations of $f(Q)$ gravity, which can be written as follows \cite{D1}
\begin{widetext}
\begin{equation}
\begin{aligned}
    \frac{-2}{\sqrt{-g}} \nabla_\alpha\left(\sqrt{-g} f_Q P_{\mu \nu}^\alpha\right)-\frac{1}{2} g_{\mu \nu} f-f_Q\left(P_{\mu \alpha \beta} Q_\nu^{\alpha \beta}-2 Q_{\alpha \beta \mu} P_\nu^{\alpha \beta}\right)=\kappa T_{\mu \nu}
\end{aligned}
\end{equation}
\end{widetext} 

Where the matter energy-momentum tensor is defined as

\begin{equation}
T_{\mu \nu}=-\frac{2}{\sqrt{-g}} \frac{\delta\left(\sqrt{-g} \mathcal{L}_m\right)}{\delta g^{\mu \nu}}  
\end{equation} 

By varying the action with respect to the affine connection, we obtain the following result \cite{ref94}
 
\begin{equation}
\nabla^\mu \nabla^\nu\left(\sqrt{-g} f_Q P^\gamma{ }_{\mu \nu}\right)=0.    
\end{equation}

The field equations in $f(Q)$ gravity preserve the conservation of the energy-momentum tensor and reduce to Einstein's equations for $f(Q)=Q$.

To apply the f(Q) gravity in the cosmology setup, we consider a spatially flat, homogeneous, and isotropic Universe described by the Friedmann-Lemaître-Robertson-Walker (FLRW) metric

\begin{equation}
ds^2 = -dt^2 + a^2(t)\left(dx^2 + dy^2 + dz^2\right),
\end{equation}
where $t$ denotes the cosmic time and $a(t)$ is the scale factor.
The Hubble parameter is defined as

\begin{equation}
H(t) \equiv \frac{\dot{a}}{a},
\end{equation}
and the cosmological redshift is related to the scale factor through
$1+z = a^{-1}$.

Within the framework of symmetric teleparallel gravity, the non-metricity scalar associated with the FLRW geometry takes the simple form

\begin{equation}
Q = 6H^2.
\end{equation}

This relation allows the gravitational sector to be fully expressed in terms of the Hubble parameter and its derivatives, significantly simplifying the cosmological equations.

Furthermore, to derive the corresponding  Friedmann equations, we assume that the cosmic matter content can be modeled as a perfect fluid, whose energy-momentum tensor is given by
\begin{equation}
T_{\mu\nu} = (\rho + p)u_\mu u_\nu + pg_{\mu\nu},
\end{equation}
where $\rho$ and $p$ denote the total energy density and pressure, respectively, and $u^\mu$ is the four-velocity of the fluid satisfying $u^\mu u_\mu = -1$.

To describe deviations from general relativity, we write the gravitational Lagrangian in the form
\begin{equation}
f(Q) = Q + F(Q),
\end{equation}
where the function $F(Q)$ encodes the modified gravity contributions.
By substituting the FLRW metric into the field equations of $f(Q)$ gravity, one obtains the generalized Friedmann equations
\begin{equation}
3H^2 = \rho + \frac{F}{2} - QF_Q,
\end{equation}
\begin{equation}
\left(2QF_{QQ} + F_Q + 1\right)\dot{H}
+ \frac{1}{4}\left(Q + 2QF_Q - F\right) = -2p,
\end{equation}
where $F_Q = dF/dQ$ and $F_{QQ} = d^2F/dQ^2$.

The total energy density is decomposed as $\rho = \rho_m + \rho_r$, corresponding to pressureless matter and radiation, respectively. The total pressure is $p = p_m + p_r$, with the matter pressure being negligible. Each matter component satisfies the standard conservation equation
\begin{equation}
\dot{\rho} + 3H(1+\omega)\rho = 0,
\end{equation}
where $\omega$ is the equation of state parameter of the corresponding fluid.

Using the explicit expression of the non-metricity scalar, the following relations are obtained
\begin{equation}
Q = 6H^2, 
\end{equation}

These relations establish a direct link between the non-metricity scalar and the cosmological dynamics, and they play a key role in determining the effective energy density and pressure arising from the modified gravity sector.

\section{Logarithmic Dark Energy Equation of State}
\label{sec:3}

In this section, we investigate the dynamical consequences of adopting a logarithmic equation of state (EoS) for the dark energy component within the framework of the power law $f(Q)$ gravity model. Unlike constant or linear parametrizations, we consider the logarithmic form \cite{ref63}

\begin{equation}
\omega_{de}(z)=\omega_0+\omega_1 \ln(1+z). \label{eq5}
\end{equation}

For a spatially flat FLRW Universe, the modified Friedmann equations take the form

\begin{align}
3H^2 &= \rho_r + \rho_m + \rho_{de}, \label{eq3} \\
2\dot H + 3H^2 &= -\frac13\rho_r - p_{de}. \label{eqX}
\end{align}

where $\rho_r$ and $\rho_m$ represent the energy densities of radiation and matter, respectively. In contrast, $\rho_{\rm de}$ and $p_{\rm de}$ describe the effective dark energy density and pressure that originate from the geometric contribution of the $f(Q)$ model

\begin{align}
\rho_{de} &= \frac{F}{2} - Q F_Q, \label{eq23} \\
p_{de} &= 2\dot H (2QF_{QQ}+F_Q) - \rho_{de}. \label{eq24}
\end{align}

Assuming no interaction among matter, radiation and dark energy,
the conservation equations are

\begin{align}
\dot\rho_r + 4H\rho_r &= 0, \label{eq1} \\  
\dot\rho_m + 3H\rho_m &= 0, \label{eq2} \\
\dot\rho_{de} + 3H(1+\omega_{de})\rho_{de} &= 0. \label{eq28}
\end{align}

The dark energy equation of state $\omega_{\rm de}$ can be expressed in terms of the function $F(Q)$ and its derivatives as

\begin{equation}
\omega_{de} = -1 +
\frac{4 \dot H ( 2QF_{QQ} + F_Q )}{F - 2QF_Q}. \label{eq29}
\end{equation}

From Eqs.~(\ref{eq1}) and ~(\ref{eq2}), the evolution of pressureless matter and radiation follows immediately, yielding the usual scalings $\rho_m \propto a^{-3}(t)$ and $\rho_r \propto a^{-4}(t)$.

\subsection{Power Law Model for $F(Q)$}

Now, we consider a specific functional form of $F(Q)$, chosen for its analytical simplicity and its ability to capture deviations from General Relativity, in a power law form
\begin{equation}
F(Q)=6\gamma H_0^2\left(\frac{Q}{Q_0}\right)^n, \label{eq31}
\end{equation}
where $H_0$, $\gamma$, $n$, and $Q_0$ are constants. This choice is motivated by the fact that Friedmann equations form a system of ordinary differential equations, for which power-law and exponential solutions can be obtained. Therefore, we adopt the power-law form in our study. Using this functional form of $f(Q)$, we first obtain

\begin{align}
\rho_{de}(z) &= 6\gamma H_0^2 (1-2n)
\left(\frac{H}{H_0}\right)^{2n}, \label{eq32} \\
\dot\rho_{de} &=
12n\gamma H_0^2 (1-2n)
\left(\frac{H}{H_0}\right)^{2n} \frac{\dot H}{H}. \label{eq33}
\end{align}

By substituting Eqs.~(\ref{eq32}) and (\ref{eq33}) into the dark energy conservation (\ref{eq28}), we obtain the key relation, together with the parametrization of the dark energy equation of state parameter. Taking into account the relation between $H$ and $Q$, we obtain 

\begin{equation}
2n \frac{\dot H}{H} + 3H(1+\omega_{de}) = 0. \label{eq34}
\end{equation}

Furthermore, using the relation between the redshift $z$ and the scale factor, namely $a(t)=1/(1+z)$, the derivative with respect to cosmic time can be rewritten in terms of the redshift variable as $\frac{d}{dt} = - (1+z) H(z) \frac{d}{dz}$. Hence, we rewrite Eq.~(\ref{eq34}) as a $z$ differential equation
\begin{equation}
-n(1+z)\frac{d H^2}{dz}
+ 3\left[1+\omega_{de}(z)\right] H^2 = 0. \label{eq36prime}
\end{equation}

\subsection{Logarithmic EoS }

By taking into account the logarithmic parametrization, Eq.~(\ref{eq5}), in Eq.~(\ref{eq36prime}), we obtain

\begin{equation}
-n(1+z)\frac{dH^2}{dz}
+ 3\left[1+\omega_0+\omega_1\ln(1+z)\right] H^2 = 0
\label{eq36log}
\end{equation}
i.e.
\begin{equation}
\frac{dH^2}{H^2}
= \frac{3}{n}\frac{1+\omega_0+\omega_1\ln(1+z)}{1+z}dz.
\end{equation}

The general solution can be written as
\begin{equation}
H^2(z)=H_0^2\,
(1+z)^{\frac{3(1+\omega_0)}{n}}
\exp\!\left[\frac{3\omega_1}{2n}\ln^2(1+z)\right],
\label{eq38prime}
\end{equation}
where $H_0$ is the present day Hubble rate. Using Eq.~(\ref{eq38prime}) in Eq.(\ref{eq32}), the dark energy density $\rho_{\rm de}$ can be expressed as
\begin{equation}
\rho_{de}(z)
= 3\gamma(1-2n)H_0^2\,
(1+z)^{3(1+\omega_0)}
\exp\!\left[\frac{3\omega_1}{2}\ln^2(1+z)\right].
\label{eq39prime}
\end{equation}

The matter density, $\rho_m$, and the radiation density, $\rho_r$, can also be expressed in terms of the redshift $z$

\begin{equation}
\rho_m(z)=\rho_{m0}(1+z)^3,
\qquad
\rho_r(z)=\rho_{r0}(1+z)^4,
\end{equation}

Thus, the Friedmann equation, Eq.~(\ref{eq3}), becomes
\begin{widetext}
\begin{equation}
\begin{aligned}
\frac{H^2(z)}{H_0^2}
\;=\; \Omega_{r0}(1+z)^4 \;+\; \Omega_{m0}(1+z)^3
\;+\; \gamma(1-2n)\,(1+z)^{3(1+\omega_0)}
\exp\!\Big(\frac{3\omega_1}{2}\ln^2(1+z)\Big).  
\end{aligned}
\end{equation}
\end{widetext}

Once the Hubble function is determined, the full background dynamics follow straightforwardly. Both matter and radiation preserve their standard scaling behaviours, while the geometrical dark energy density inherits the same logarithmic modulation present in the EoS. The resulting Friedmann equation provides the basis for the observational analysis conducted in the subsequent section.

For consistency with standard cosmological analyses, we adopt the normalization
\begin{equation}
H = 100\,h \;\mathrm{km\,s^{-1}\,Mpc^{-1}}.
\end{equation}
In this framework, the parameter $\gamma$ is treated as a derived quantity, calculated from the matter density $\Omega_{m0}$ and the power law index $n$ via the relation
\begin{equation}
\gamma(\Omega_{m0}, n)= \frac{1 - \Omega_{m0} - \Omega_{r0}}{1 - 2 n},
\end{equation}
where $\Omega_{r0}$ denotes the present day radiation density parameter, which is fixed to $\Omega_{r0} = 8.4 \times 10^{-5}$. This choice ensures that the model reduces to the correctly normalized Hubble rate at the present epoch.

\section{Observational Data and Statistical Methodology}
\label{sec:4}

\subsection{Methodology}

In this subsection, we briefly describe the Markov Chain Monte Carlo (MCMC) methodology employed to constrain the free parameters of the power law $f(Q)$ model. The MCMC technique is widely used in cosmology to efficiently explore high dimensional parameter spaces and to derive robust posterior probability distributions for cosmological parameters \cite{ref64}.

The fundamental idea of the MCMC approach is to generate a Markov chain that samples the parameter space according to a target probability distribution. Each point in the chain is proposed based on the previous one through a transition probability governed by a chosen proposal distribution, ensuring convergence toward the posterior distribution \cite{ref72,ref73}.

In our analysis, We constrain the model parameters using three independent observational datasets: the Pantheon$^{+}$ Type Ia supernova compilation, consisting of 1701 data points~\cite{ref66}; the DESI DR2 BAO dataset, which includes 12 data points~\cite{AbdulKarim2025}; and the Cosmic Chronometer (CC) measurements, comprising 32 data points~\cite{Moresco2020}. These datasets provide complementary constraints on the expansion history of the Universe.

The best-fit values of the model parameters are obtained by maximizing the likelihood function, which is assumed to have a Gaussian form. The joint analysis of multiple datasets allows us to derive tightly constrained posterior distributions for the parameters h, $\Omega_{m0}$, and n, as well as for the effective dark energy equation of state parameters $\omega_0$ and $\omega_1$.
\begin{equation}
\mathcal{L} \propto \exp\Biggl(-\frac{\chi^2}{2}\Biggr),
\end{equation}
where $\chi^2$ denotes the pseudo chi-square estimator \cite{ref64}.
For independent datasets, the total likelihood is given by the product of the individual likelihoods, which leads to a total chi-square

\begin{equation}
\chi^2_{\mathrm{tot}} = \chi^2_{\mathrm{CC}} + \chi^2_{\rm Pantheon^+} + \chi^2_{\rm DESI\ DR2}
\end{equation}

The explicit construction of the $\chi^2$ functions associated with each dataset is presented in the following subsections. The resulting MCMC chains are then used to derive the marginalized posterior distributions and confidence intervals for all model parameters.

\subsection{Datasets}
\subsubsection{Cosmic Chronometers}
We use a compilation of 32 cosmic chronometer (CC) measurements of the Hubble parameter $H(z)$ in the redshift range $0 < z \lesssim 2$. The CC method provides a direct and model-independent estimate of the expansion rate based on the differential age evolution of passively evolving galaxies, through the relation
\begin{equation}
H(z) = -\frac{1}{1+z} \frac{dz}{dt}.
\end{equation}
We adopt the dataset of Moresco et al.~\cite{Moresco2020}. including the full covariance matrix to account for correlated systematic uncertainties\footnote{The covariance matrix used in this work is publicly available at \url{https://github.com/Ahmadmehrabi/Cosmic_chronometer_data}.}, such as those arising from metallicity, young stellar population contamination, and stellar population synthesis modeling. The $\chi^2$ function for the cosmic chronometer data is defined as
\begin{equation}
\chi^2_{\mathrm{CC}} = \Delta \mathbf{H}^T \, \mathbf{C}^{-1} \, \Delta \mathbf{H},
\end{equation}
where $\Delta \mathbf{H} = H_{\mathrm{th}}(z_i) - H_{\mathrm{obs}}(z_i)$, with $H_{\mathrm{obs}}(z_i)$ and $H_{\mathrm{th}}(z_i)$ denoting the observed and theoretical values of the Hubble parameter at redshift $z_i$, respectively. The covariance matrix $\mathbf{C}$ includes both statistical and systematic contributions.
\subsubsection{Pantheon$^{+}$ dataset}
We include the Pantheon$^{+}$ compilation of Type Ia supernovae~\cite{ref66}. It comprises 1701 data points obtained from 1550 supernovae covering the redshift range $0.001 \leq z \leq 2.3$. The chi-square function associated with the supernova dataset is given by
\begin{equation}
\chi_{\text{Pantheon}^{+}}^2
= \vec{F}^{\,T} \cdot \mathbf{C}_{\text{Pantheon}^{+}}^{-1} \cdot \vec{F},
\label{eq11}
\end{equation}
where $C_{\rm Pantheon^{+}}$ denotes the covariance matrix derived from the Pantheon$^{+}$ dataset, incorporating both statistical and systematic uncertainties. The parameter vector is defined as $\theta = (h, \Omega_{m0}, \omega_0, \omega_1, n, M)$, where $M$ represents the absolute magnitude. In this expression,
\begin{equation}
\vec{F}_i = m_{B,i} - M - \mu_{\rm model},
\end{equation}
where $m_{B,i}$ and $\mu_{\rm model}$ represent the observed apparent magnitudes and the theoretical distance modulus, respectively. The predicted distance modulus, $\mu_{\rm model}$, is computed according to the chosen cosmological model as follows
 \begin{equation}
\mu_{\rm model}(z) = m - M = 5 \log_{10} D_L(z) + 25,
\end{equation}
where $D_L$ represents the luminosity distance, defined as
\begin{equation}
D_L(z) = (1+z) \int_0^z \frac{dz^\ast}{H(z^\ast)},
\end{equation}
where, $c$ denotes the speed of light. Unlike the original Pantheon dataset, the Pantheon+ compilation breaks the degeneracy between the absolute magnitude $M$ and the Hubble constant $H_0$. This is achieved by expressing the vector $\vec{F}$ in Eq.~(\ref{eq11}) in terms of the distance modulus of Type Ia supernovae hosted in Cepheid galaxies. Consequently, an independent constraint on $M$ is obtained, leading to the following expression

\begin{equation}
\vec{F}_i^{\prime} = 
\begin{cases} 
m_{B,i} - M - \mu_i^{\rm Ceph} & i \in \text{Cepheid hosts} \\[1em]
m_{B,i} - M - \mu_{\rm model}(z_i) & \text{otherwise} 
\end{cases}
\end{equation}
where, $\mu_i^{\rm Ceph}$ denotes the distance modulus of the Cepheid host galaxy for the $i$-th Type Ia supernova, independently determined using Cepheid calibrators. Accordingly, Eq.~(\ref{eq11}) can be rewritten as

\begin{equation}
  \chi_{\rm SN}^2 = \vec{F}^{\prime \, T} \cdot \mathbf{C}_{\text{Pantheon}^{+}}^{-1} \cdot \vec{F}^{\prime}.  
\end{equation}

\subsubsection{DESI DR2 dataset}

We use the DESI DR2 Baryon Acoustic Oscillation (BAO) dataset to constrain the power law $f(Q)$ model. This dataset includes several tracers, namely Bright Galaxy Survey (BGS) galaxies, Luminous Red Galaxies (LRGs), Emission Line Galaxies (ELGs), quasars (QSO), and the Lyman-$\alpha$ forest, covering the redshift range $0.1 \leq z \leq 4.2$ \cite{AbdulKarim2025}.

From these observations, we consider the comoving angular diameter distance $D_M(z)/r_d$ and the Hubble distance $D_H(z)/r_d$, defined as 
\begin{equation}
D_M(z) \equiv \int_0^z \frac{c \, dz'}{H(z')}, \quad \text{and} \quad
D_H(z) \equiv \frac{c}{H(z)},
\end{equation}
where $r_d$ is the sound horizon at the drag epoch. Assuming standard early-Universe physics, $r_d$ can be defined as 
\begin{equation}
    r_d \equiv r_s\left(z_d\right)=\int_{z_d}^{\infty} \frac{c_s\left(z^{\prime}\right)}{H\left(z^{\prime}\right)} d z^{\prime},
\end{equation}
where $z_d$ denotes the redshift of the drag epoch and $c_s$ is the sound speed of the baryon-photon fluid. We also include the volume averaged distance
\begin{equation}
D_V(z) \equiv \left[z \, D_M^2(z) \, D_H(z)\right]^{1/3}.
\end{equation}
The DESI DR2 BAO distance measurements used in this work can be found in Table~III of \cite{C3}. For these, we use the covariance matrices provided in the DESI DR2 release.

The chi-square for the DESI BAO dataset is defined as
\begin{equation}
\chi^2_{\mathrm{DESI}} =
\Delta D^{\,T}\, \mathbf{C}^{-1}\, \Delta D,
\end{equation}
where $\Delta D = D_{\mathrm{obs}} - D_{\mathrm{th}}$ denotes the difference between observed and theoretical BAO measurements, and $\mathbf{C}$ is the covariance matrix. For uncorrelated data, this reduces to
\begin{equation}
\chi^2_{\mathrm{DESI}} = \sum_i \left(\frac{D^{\mathrm{obs}}_i - D^{\mathrm{th}}_i}{\sigma_i}\right)^2.
\end{equation}
\begin{table}
\centering
\caption{Priors for the parameter space h, $\Omega_{m0}$, $M$, $\omega_0$, $\omega_1$, $n$.}
\label{tab:2}
\begin{tabular}{|l|l|}
\hline
Parameter & Prior \\
\hline
$h$            & $(0.6,0.8)$ \\
$\Omega_{m0}$  &  $(0,1)$ \\
$M$  &  $(-20,-19)$ \\
$\omega_0$     & $(-2,0)$ \\
$\omega_1$     & $(-2,2)$ \\
$n$       & $(-2,2)$ \\
$r_d$       & $(100, 200)$ \\
\hline
\end{tabular}
\end{table}

In this work, we consider three dataset combinations: Pantheon$^{+}$ only, Pantheon$^{+}$ + DESI DR2, and Pantheon$^{+}$ + DESI DR2 + CC. The adopted prior ranges are listed in Table~\ref{tab:2}.

\section{Numerical Results and Discussion}
\label{sec:5}

Table~\ref{tab:1} summarizes the mean values and the corresponding $68\%$ confidence intervals of the cosmological parameters obtained for $\Lambda$CDM and power law $f(Q)$ models using three dataset combinations: Pantheon$^{+}$ only, Pantheon$^{+}$+DESI DR2, and Pantheon$^{+}$+DESI DR2+CC. Overall, both models lead to stable and well constrained parameter estimates, providing a consistent description of the observational data.

\begin{figure*}[t]
\centering
\includegraphics[width=1.04\textwidth]{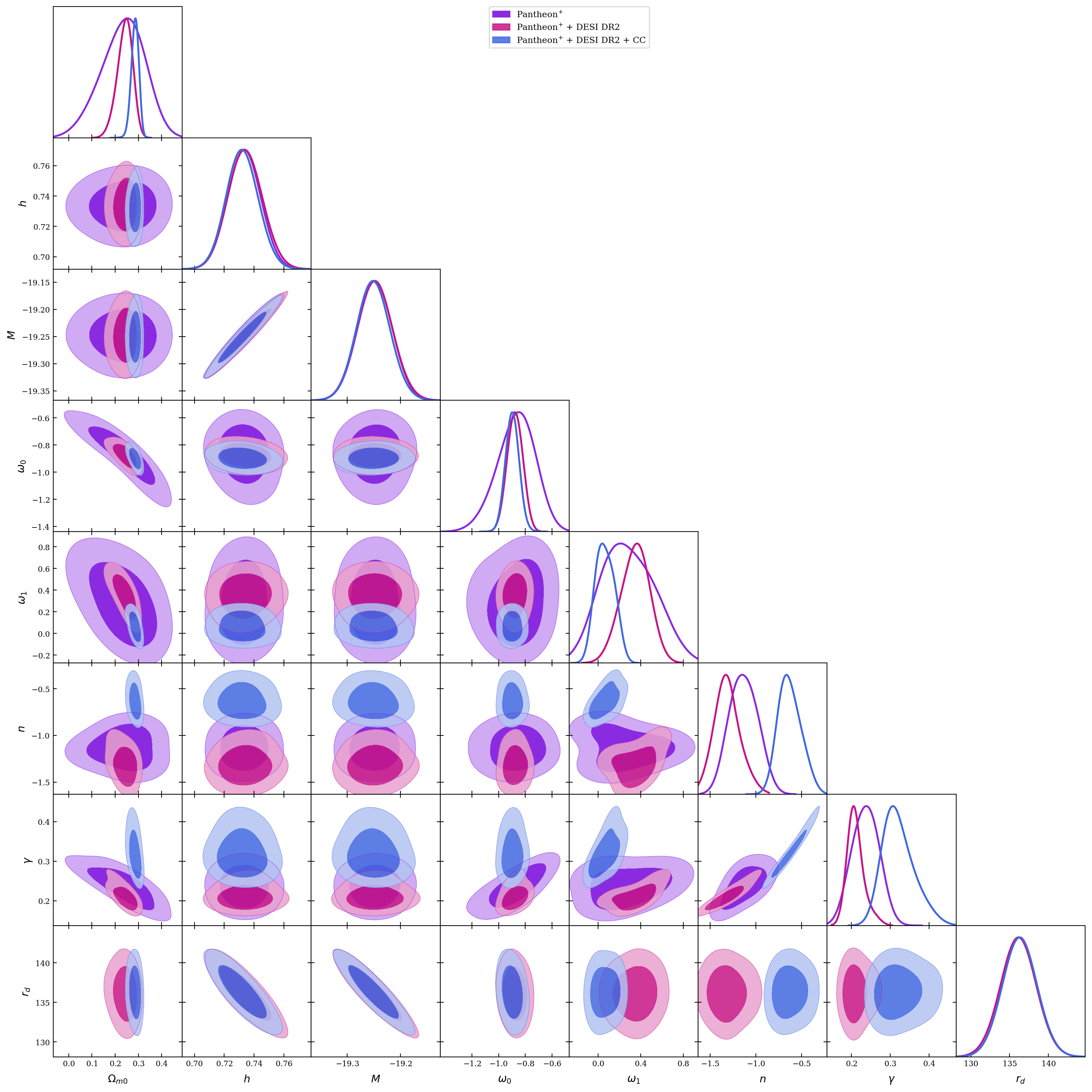}
\caption{The $1\sigma$ and $2\sigma$ confidence contours and posterior distributions for the power law $f(Q)$ model using Pantheon$^{+}$, DESI DR2, and CC datasets.}
\label{fig:5}
\end{figure*}

For the MCMC analysis of the combined datasets, we adopt the same priors as for the CC sample. The resulting $1\sigma$ and $2\sigma$ confidence contours are shown in Fig.~\ref{fig:5}.

For the Hubble parameter, the inferred values remain tightly constrained across all data combinations. Using the Pantheon$^{+}$ dataset alone, we obtain $h = 0.735 \pm 0.010$ and $h = 0.733 \pm 0.010$ for the $\Lambda$CDM and power law models, respectively, in good agreement with the SH0ES measurement ($H_0 = 73 \pm 2$ km s$^{-1}$ Mpc$^{-1}$). Including DESI DR2 data leads to consistent results, with $h = 0.736 \pm 0.011$ and $h = 0.734 \pm 0.011$ for the $\Lambda$CDM and power law models, respectively. Finally, using the combined Pantheon$^{+}$ + DESI DR2 + CC dataset, we obtain $h = 0.735 \pm 0.011$ and $h = 0.732 \pm 0.010$ for the $\Lambda$CDM and power law models, respectively. The close agreement between the two models indicates that the modified gravity framework accurately reproduces the background expansion history.

The matter density parameter $\Omega_{m0}$ is constrained around $0.334 \pm 0.019$ and $0.235^{+0.10}_{-0.081}$ using Pantheon$^{+}$ alone, $0.3070 \pm 0.0097$ and $0.242^{+0.036}_{-0.028}$ with Pantheon$^{+}$ + DESI DR2, and $0.3075 \pm 0.0099$ and $0.286^{+0.017}_{-0.015}$ for the combined Pantheon$^{+}$ + DESI DR2 + CC dataset, for the $\Lambda$CDM and power law models, respectively. This result confirms that the power law $f(Q)$ model remains consistent with the standard matter content required by the observed cosmic expansion.

In the power law $f(Q)$ scenario, the dark energy sector is described through a logarithmic parameterization of the equation of state, as seen in Eq.~(\ref{eq5}), which allows for a mild redshift dependence. The constraints reported in Table~\ref{tab:1} indicate $w_0 = -0.87^{+0.15}_{-0.13}$ and $w_1 = 0.28^{+0.23}_{-0.28}$ from Pantheon$^{+}$, $w_0 = -0.877 \pm 0.056$ and $w_1 = 0.35^{+0.14}_{-0.12}$ from Pantheon$^{+}$+DESI DR2, and $w_0 = -0.898 \pm 0.050$ and $w_1 = 0.065 \pm 0.087$ from Pantheon$^{+}$+DESI DR2+CC. These results show that the combination of datasets progressively improves the constraints on both $w_0$ and $w_1$, with the tightest bounds obtained when all data sets are included.

\begin{figure*}[t]
    \centering    
    \includegraphics[width=0.65\linewidth]{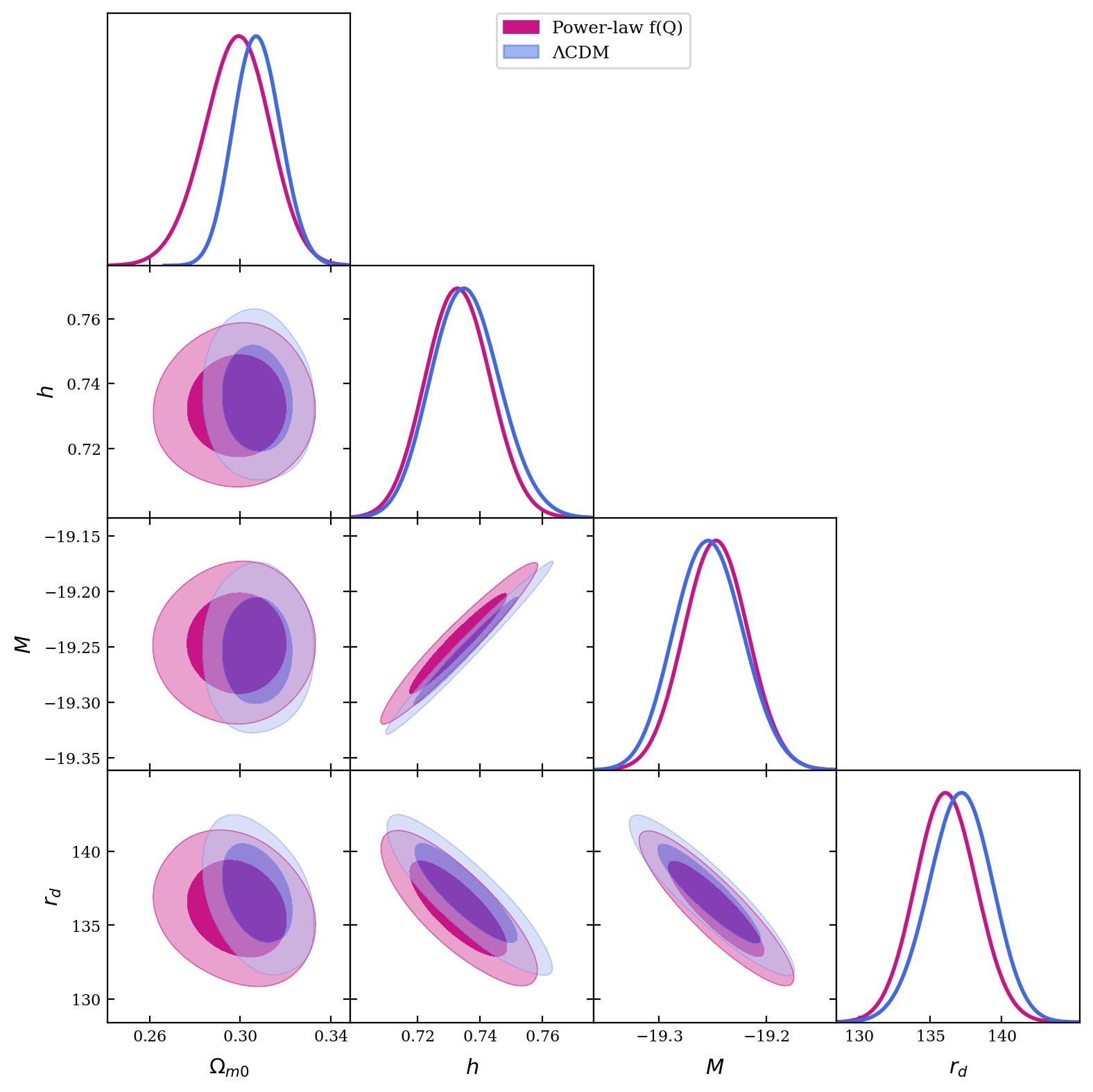}
    \caption{The $1\sigma$ and $2\sigma$ confidence contours, along with the posterior distributions, obtained for the power law $f(Q)$ and $\Lambda$CDM models using the Pantheon$^{+}$+ DESI DR2 + CC datasets.}
    \label{fig:contours1}
\end{figure*}

The posterior distributions and the joint confidence contours shown in Figs.~\ref{fig:5} and \ref{fig:contours1} further illustrate the impact of this parametrization. Fig.~\ref{fig:5} presents the marginalized confidence levels at $1\sigma$ and $2\sigma$ for Pantheon$^{+}$ alone, Pantheon$^{+}$+DESI DR2, and Pantheon$^{+}$+DESI DR2+CC. Using the joint analysis, Fig.~\ref{fig:contours1} presents the confidence contours for both the $\Lambda$CDM model and the power law $f(Q)$ model. A clear correlation between $\omega_0$ and $\omega_1$ is observed, reflecting their combined role in shaping the redshift evolution of $\omega_{\mathrm{de}}(z)$. The contours also demonstrate that the inclusion of multiple datasets significantly tightens the constraints, leading to well localized confidence regions for all parameters. In particular, the power law index and the coupling parameter are well constrained around their best-fit values, indicating that the additional degrees of freedom introduced by the model are effectively supported by the data.

\begin{table*}[t!] 
\centering
\caption{Mean values and standard deviations of the cosmological parameters for the $\Lambda$CDM and power law $f(Q)$ models.}
\label{tab:1}
\small
\begin{tabular}{lcccccccc}
\hline
Model & $h$ & $\Omega_{m0}$ & $M$ & $\omega_0$ & $\omega_1$ & $n$ & $\gamma$ & $r_d$\\
\hline
\multicolumn{8}{c}{Pantheon$^+$} \\
\hline
$\Lambda$CDM  & $0.735 \pm 0.010$ & $0.334 \pm 0.019$ & $-19.246 \pm 0.030$ & -- & -- & -- & -- & -- \\
Power law     & $0.733 \pm 0.010$ & $0.235^{+0.10}_{-0.081}$ & $-19.248 \pm 0.030$ & $-0.87^{+0.15}_{-0.13}$ & $0.28^{+0.23}_{-0.28}$ & $-1.13 \pm 0.15$ & $0.236 \pm 0.033$ & -- \\
\hline
\multicolumn{8}{c}{Pantheon$^+$ + DESI DR2} \\
\hline
$\Lambda$CDM  & $0.736 \pm 0.011$ & $0.3070 \pm 0.0097$ & $-19.252 \pm 0.031$ & -- & -- & -- & -- & $137.0 \pm 2.1$ \\
Power law     & $0.734 \pm 0.011$ & $0.242^{+0.036}_{-0.028}$ & $-19.248 \pm 0.031$ & $-0.877 \pm 0.056$ & $0.35^{+0.14}_{-0.12}$ & $-1.32^{+0.12}_{-0.15}$ & $0.210^{+0.015}_{-0.023}$ & $136.1 \pm 2.2$ \\
\hline
\multicolumn{8}{c}{Pantheon$^+$ + DESI DR2 + CC} \\
\hline
$\Lambda$CDM  & $0.735 \pm 0.011$ & $0.3075 \pm 0.0099$ & $-19.253 \pm 0.031$ & -- & -- & -- & -- & $137.1 \pm 2.1$ \\
Power law     & $0.732 \pm 0.010$ & $0.286^{+0.017}_{-0.015}$ & $-19.251 \pm 0.030$ & $-0.898 \pm 0.050$ & $0.065 \pm 0.087$ & $-0.63^{+0.11}_{-0.14}$ & $0.320^{+0.033}_{-0.047}$ & $136.3 \pm 2.2$ \\
\hline
\end{tabular}
\end{table*}

\subsection{Information criteria}

Since the cosmological models considered in this work involve different numbers of free parameters, the minimum chi-square statistic $\chi^2_{\min}$ alone is not sufficient to provide a reliable model comparison. To assess the relative performance of models based on their theoretical predictions and their consistency with observational data, we employ three commonly used statistical criteria: the reduced chi-square $\chi^2_{\rm red}$, the corrected Akaike information criterion (AIC$_c$)~\cite{ref70}, and the Bayesian information criterion (BIC)~\cite{ref71}. These quantities are defined as
\begin{equation}
\chi^2_{\rm red} = \frac{-2 \ln \mathcal{L}_{\max}}{N_d - N_p},
\end{equation}
\begin{equation}
\mathrm{AIC_c} = -2 \ln \mathcal{L}_{\max} + 2 N_p
+ \frac{2 N_p (N_p + 1)}{N_d - N_p - 1},
\end{equation}
and
\begin{equation}
\mathrm{BIC} = -2 \ln \mathcal{L}_{\max} + N_p \ln (N_d),
\end{equation}
where $\mathcal{L}_{\max}$ denotes the maximum likelihood, $N_d$ represents the total number of data points, and $N_p$ is the number of free parameters of the model. The AIC and BIC values, together with their differences relative to the reference $\Lambda$CDM model, are reported in Table~\ref{tab:3} for the considered datasets.

To compare the statistical performance of the models, we consider the differences of the information criteria with respect to the reference $\Lambda$CDM model, defined as $\Delta\mathrm{AIC} = \mathrm{AIC}_{\text{model}} - \mathrm{AIC}_{\Lambda\mathrm{CDM}}$ and $\Delta\mathrm{BIC} = \mathrm{BIC}_{\text{model}} - \mathrm{BIC}_{\Lambda\mathrm{CDM}}$. In this framework, negative values indicate that the considered model is statistically favored over $\Lambda$CDM, while positive values imply a preference for the reference model. Typically, $\Delta\mathrm{AIC} \lesssim 2$ suggests that the two models are statistically indistinguishable, whereas larger values indicate decreasing support for the alternative model. For the BIC criterion, $\Delta\mathrm{BIC} > 6$ is commonly interpreted as strong evidence against the model under consideration.

For the Pantheon$^+$ + DESI DR2 dataset, the power law $f(Q)$ model gives $\Delta\mathrm{AIC} = 0.91$, indicating that both models are statistically comparable, with a slight preference for $\Lambda$CDM. For the Pantheon$^+$ + DESI DR2 + CC dataset, we find $\Delta\mathrm{AIC} = 1.06$, which also suggests no significant preference between the two models according to the AIC criterion.

However, the BIC yields $\Delta\mathrm{BIC} = 17.26$ and $17.44$ for Pantheon$^+$ + DESI DR2 and Pantheon$^+$ + DESI DR2 + CC, respectively. These values indicate a strong penalty on the power law $f(Q)$ model due to its larger number of free parameters, thus clearly favoring the $\Lambda$CDM model.

The Pantheon$^+$+DESI DR2 dataset shows comparable performance between the $\Lambda$CDM and power law $f(Q)$ models according to the AIC criterion, while the addition of CC data further tightens the constraints. Taken together, these results demonstrate that the power law $f(Q)$ model provides a statistically consistent and phenomenologically viable description of the observational data. The logarithmic dark energy parameterization allows for a controlled departure from a strictly constant equation of state, while remaining closely aligned with the $\Lambda$CDM scenario within current observational constraints.

\begin{table*}[t!]
\centering
\caption{The corresponding $\chi^2_{\min}$, AICc and BIC for the examined cosmological models.}
\label{tab:3}
\small
\begin{tabular}{lcccccc}
\hline
Model & $\chi^2_{\min}$ & $\chi^2_{\rm red}$ & AIC & $\Delta$AIC & BIC & $\Delta$BIC  \\
\hline
\multicolumn{7}{c}{Pantheon$^+$ + DESI DR2} \\
\hline
$\Lambda$CDM  & 1536.31 & 0.899 & 1544.31 & 0 & 1566.11 & 0 \\
Power law     & 1531.22 & 0.897 & 1545.22 & 0.91 & 1583.37 & 17.26  \\
\hline
\multicolumn{7}{c}{Pantheon$^+$ + DESI DR2 + CC} \\
\hline
$\Lambda$CDM  & 1546.71 & 0.888 & 1554.71 & 0 & 1576.55 & 0  \\
Power law     & 1541.77 & 0.887 & 1555.77 & 1.06 & 1593.99 & 17.44  \\
\hline
\end{tabular}
\end{table*}

\section{Cosmographic and Dynamical Diagnostics}
\label{sec:6}

Cosmography provides a model-independent framework to describe the expansion history of the Universe, relying only on the assumptions of homogeneity and isotropy. This approach allows the kinematical behavior of cosmic expansion to be characterized without specifying a particular gravitational theory or dark energy model, making it a valuable tool for confronting theoretical predictions with observational data.

The aim of this section is to investigate a cosmological model capable of describing the global dynamics of the Universe through its large-scale physical and geometrical properties. It is convenient to expand the scale factor $a(t)$ as a Taylor series around the present cosmic time $t_0$
\begin{equation}
\begin{aligned}
\frac{a(t)}{a_0} = &\, 1 + \left.\frac{(t - t_0)}{1!}\frac{da}{dt}\right|_{t_0}
+ \left.\frac{(t - t_0)^2}{2!}\frac{d^2 a}{dt^2}\right|_{t_0} \\
& + \frac{(t - t_0)^3}{3!}\frac{d^3 a}{dt^3}
+ \frac{(t - t_0)^4}{4!}\frac{d^4 a}{dt^4} + \cdots
\end{aligned}
\end{equation}
where $a_0$ denotes the present value of the scale factor.

The coefficients appearing in this expansion can be expressed in terms of cosmographic parameters defined as
\begin{equation}
\begin{aligned}
H &= \frac{1}{a}\frac{da}{dt}, \qquad
q(t) = -\frac{1}{aH^2}\frac{d^2 a}{dt^2}, \\
j(t) &= \frac{1}{aH^3}\frac{d^3 a}{dt^3}, \qquad
s(t) = \frac{1}{aH^4}\frac{d^4 a}{dt^4},
\end{aligned}
\end{equation}
which are commonly known as the Hubble, deceleration, jerk, and snap parameters. These quantities provide a purely kinematical description of the cosmic expansion and are sufficient to characterize the overall behavior of the Universe.

For practical purposes, it is useful to rewrite these parameters in terms of the redshift $z$ \cite{C1,C2}. They can then be expressed as
\begin{equation}
\begin{aligned}
q(z) &= -1 + (1+z)\frac{E'(z)}{E(z)}, \\
j(z) &= (1+z)^2 \frac{E''(z)}{E(z)} + q^2(z), \\
s(z) &= -(1+z) j'(z) - 2 j(z) - 3 q(z) j(z),
\end{aligned}
\end{equation}
where $E(z)=\frac{H(z)}{H_0}$ denotes the normalized Hubble parameter.
\subsection{Deceleration parameter}
\begin{figure}[H]
    \centering
    \includegraphics[width=1.1\linewidth]{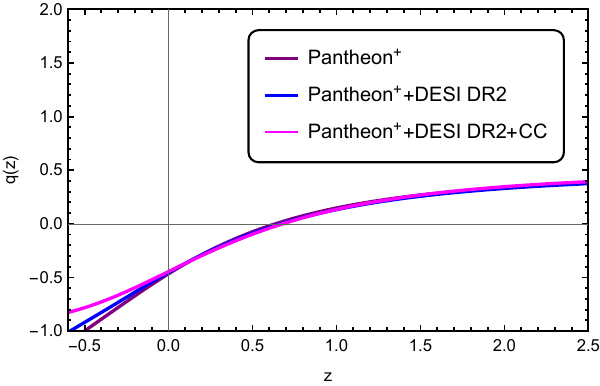}
    \caption{Evolution of the deceleration parameter as a function of redshift $z$.}
    \label{fig:placeholder}
\end{figure}

The deceleration parameter, shown in Fig.~\ref{fig:placeholder}, exhibits a smooth transition from an early decelerated phase, associated with matter domination, to a late-time accelerated expansion. A positive value of $q$ corresponds to a decelerating Universe, while a negative value indicates an accelerating expansion. This behavior reflects the transition from deceleration to acceleration in the recent past, with transition redshifts $z_t \approx 0.635, 0.666, 0.679$ for the Pantheon$^{+}$, Pantheon$^{+}$+DESI DR2, and Pantheon$^{+}$+DESI DR2+CC datasets, respectively. At the present epoch, $q_0 \approx -0.445, -0.455, -0.457$ for the same datasets, confirming the current accelerated expansion. The close agreement among these values underscores the robustness of the model with respect to the inclusion of additional data, while future predictions are sensitive to the data.
\subsection{Jerk parameter}
\begin{figure}[H]
    \centering
    \includegraphics[width=1.1\linewidth]{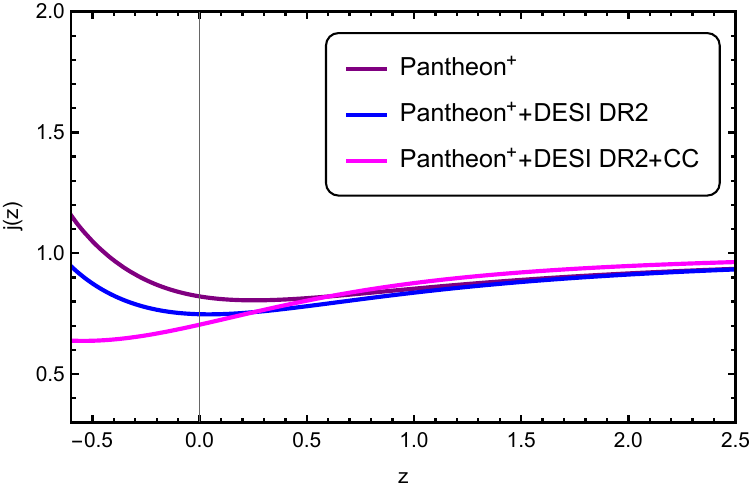}
    \caption{Evolution of the jerk parameter $j$ as a function of redshift $z$.}
    \label{fig:6}
\end{figure}

The jerk parameter, shown in Fig.~\ref{fig:6}, evolves smoothly over the entire redshift range, with no indication of pathological behavior. At the present epoch, the values $j_0 \approx 0.81, 0.74, 0.70$ for the Pantheon$^{+}$, Pantheon$^{+}$+DESI DR2, and Pantheon$^{+}$+DESI DR2+CC datasets, respectively, remain close to the $\Lambda$CDM reference value $j= 1$. The slight decrease of $j_0$ with the inclusion of additional datasets reflects tighter observational constraints, while preserving values near unity. This behavior indicates that the model reproduces the main kinematic features of the standard cosmological scenario with only mild deviations at late times. In the recent past and at earlier epochs, the evolution is not sensitive to the data, whereas future predictions become more sensitive to the choice and precision of the observational datasets.
\subsection{Snap parameter}
\begin{figure}[H]
    \centering
    \includegraphics[width=1.1\linewidth]{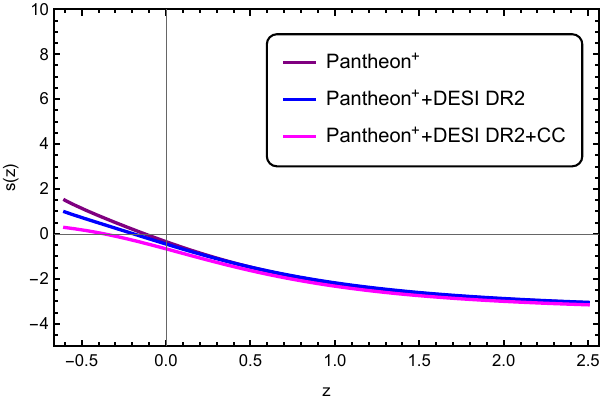}
    \caption{Evolution of the snap parameter $s$ as a function of redshift $z$.}
    \label{fig:7}
\end{figure}

The evolution of the snap parameter $s(z)$ for the present model is illustrated in Fig.~\ref{fig:7}. This parameter captures higher order variations of the cosmic expansion and evolves in a smooth and regular manner across the full redshift range, indicating the stability of the model at higher derivatives. At the present epoch, $s_0$ takes the values $-0.37$ for the Pantheon$^{+}$ dataset, $-0.46$ for Pantheon$^{+}$+DESI DR2, and $-0.65$ when CC data are included, highlighting the impact of additional observational constraints on higher order aspects of the expansion history. Nevertheless, its overall behavior remains well controlled and consistent with a smooth late-time expansion, confirming the robustness and reliability of the model in describing the recent evolution of the Universe. In the recent past and at earlier epochs, the evolution remains insensitive to the observational data, while future predictions become increasingly sensitive to the quality and precision of the datasets.

\subsection{EoS Parameter}

\begin{figure}[H]
    \centering
    \includegraphics[width=1.1\linewidth]{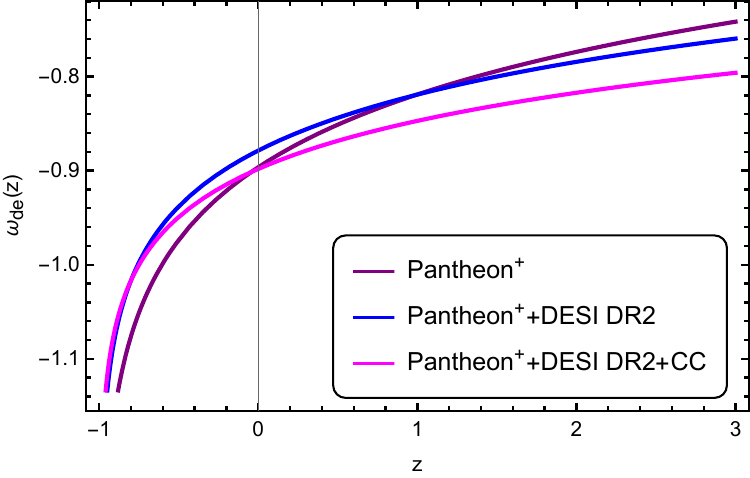}
    \caption{Evolution of the dark energy equation of state $\omega_{de}$ as a function of redshift $z$.}
    \label{fig:EoS}
\end{figure}

The evolution of the dark energy equation of state $\omega_{de}(z)$, shown in Fig.~\ref{fig:EoS}, exhibits a smooth and mild redshift dependence. At the present epoch ($z=0$), the values are $\omega_{de} = -0.897$ for Pantheon$^{+}$, $\omega_{de} = -0.878$ for Pantheon$^{+}$+DESI DR2, and $\omega_{de} = -0.898$ for Pantheon$^{+}$+DESI DR2+CC, indicating a quintessence-like behavior.

A crossing of the phantom divide line ($\omega_{de} = -1$) occurs at negative redshift, with $z \simeq -0.604$ for Pantheon$^{+}$, $z \simeq -0.756$ for Pantheon$^{+}$+DESI DR2, and $z \simeq -0.748$ for Pantheon$^{+}$+DESI DR2+CC. This indicates that the transition to the phantom regime is shifted further into the future when additional datasets are included.

\subsection{Density parameters}

\begin{figure}[H]
    \centering
    \includegraphics[width=1.1\linewidth]{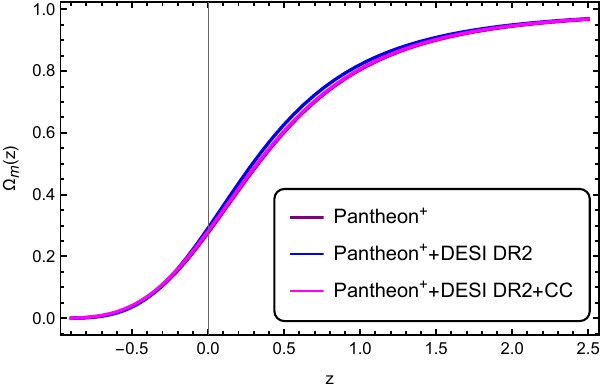}
    \caption{Evolution of the matter energy density parameter $\Omega_m$ as a function of redshift $z$.}
    \label{fig:8}
\end{figure}

\begin{figure}[H]
    \centering
    \includegraphics[width=1.1\linewidth]{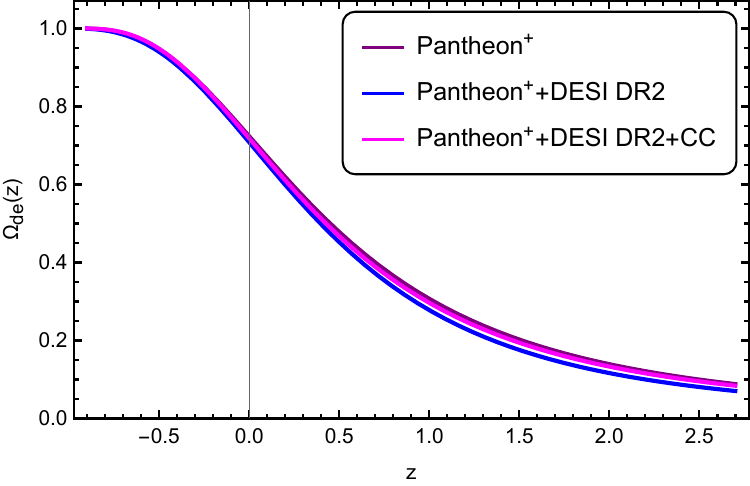}
    \caption{Evolution of the dark energy density parameter $\Omega_{\rm de}$ as a function of redshift $z$.} 
    \label{fig:10}
\end{figure}

The evolution of the matter and dark energy density parameters, $\Omega_m(z)$ and $\Omega_{\rm de}(z)$, is illustrated in Figs.~\ref{fig:8} and \ref{fig:10} for the different observational datasets. As expected, $\Omega_m(z)$ increases with redshift, indicating that matter dominated the Universe in the recent past, while $\Omega_{\rm de}(z)$ decreases with redshift, showing that dark energy only becomes dominant at late times. This complementary behavior reflects a smooth transition from a matter-dominated era to a dark energy-dominated phase driving the current accelerated expansion.  

At the present epoch, the matter density parameter takes the values $\Omega_{m0} = 0.279$, $0.290$, and $0.288$ for Pantheon$^{+}$, Pantheon$^{+}$+DESI DR2, and Pantheon$^{+}$+DESI DR2+CC, respectively, while the corresponding dark energy density parameter values are $\Omega_{{\rm de}0} = 0.720$, $0.708$, and $0.717$. These results confirm that dark energy is the dominant component at low redshift, with matter contributing a subdominant but non negligible fraction of the total energy density.  

In addition, the matter density parameter $\Omega_m(z)$ is insensitive to the observational datasets. In contrast, the dark energy density parameter $\Omega_{\rm de}(z)$ exhibits a slightly higher level of sensitivity to the choice of dataset, reflecting the mild dependence of the dark energy sector on observational constraints. Nevertheless, these variations remain small and do not affect the overall evolutionary behavior.

Overall, despite minor differences induced by the inclusion of additional observational data, the evolution of both components remains consistent, with dark energy expected to continue dominating the cosmic energy budget in the near future according to the present model.

In terms of the Hubble parameter, the density parameters are defined as
\begin{equation}
\Omega_m(z) = \frac{\rho_{m0}\,(1+z)^3}{3H^2(z)},
\end{equation}
and
\begin{equation}
\Omega_{\rm de}(z) = \frac{\rho_{{\rm de}0}\,(1+z)^{3(1+\omega_{\rm de})}}{3H^2(z)},
\end{equation}
where $\rho_{m0}$ and $\rho_{{\rm de}0}$ denote the present day matter and dark energy densities, respectively, and $\omega_{\rm de}$ represents the effective dark energy equation of state.

\subsection{Om diagnostic}

\begin{figure}
    \centering
    \includegraphics[width=1.1\linewidth]{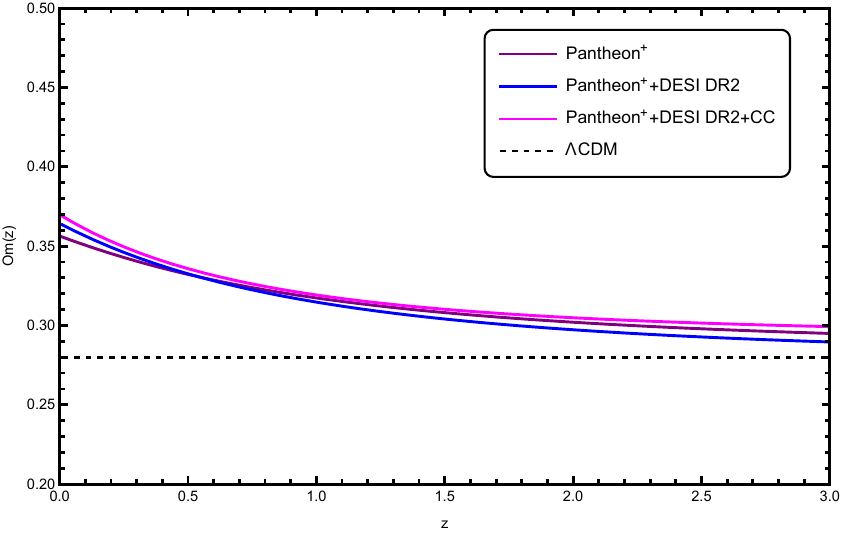}
    \caption{Evolution of the $\mathrm{Om}(z)$ Diagnostic as a function of redshift $z$.}
    \label{fig:11}
\end{figure}

The $\mathrm{Om}(z)$ diagnostic provides a simple and robust test to distinguish the standard $\Lambda$CDM model from scenarios with dynamical dark energy, as it depends only on the Hubble parameter and is less sensitive to observational uncertainties. For a spatially flat Universe, it is defined as
\begin{equation}
   \mathrm{Om}(z) = \frac{E^2(z)-1}{(1+z)^3-1}. 
\end{equation} 
As illustrated in Fig.~\ref{fig:11}, the reconstructed $\mathrm{Om}(z)$ curves for the Pantheon$^{+}$, Pantheon$^{+}$+DESI DR2, and Pantheon$^{+}$+DESI DR2+CC datasets lie below the constant $\Lambda$CDM value, indicating a deviation from a strictly constant dark energy component and suggesting an effective phantom-like behavior. Moreover, the negative slope of the $\mathrm{Om}(z)$ curves provides additional evidence for a phantom regime. The combination of multiple datasets, namely Pantheon$^{+}$+DESI DR2+CC, leads to tighter constraints and a smoother evolution. Overall, the $\mathrm{Om}(z)$ diagnostic supports the presence of a dynamical dark energy component within the framework of power law $f(Q)$ gravity.

\subsection{Statefinder diagnostic}

The statefinder diagnostic offers a geometrical approach for discriminating among different dark energy models by involving higher order derivatives of the scale factor. It is characterized by a pair of dimensionless parameters $\{j,s\}$, which provide a model independent description of the cosmic expansion history and allow for a direct comparison between the $\Lambda$CDM model and its extensions. The statefinder parameters are defined as~\cite{ref86,ref87}
\begin{equation}
s = \frac{j-1}{3\left(q-\frac{1}{2}\right)},
\end{equation}
where $j$ is the jerk parameter and $q$ denotes the deceleration parameter.

\begin{figure}
    \centering
    \includegraphics[width=1.0\linewidth]{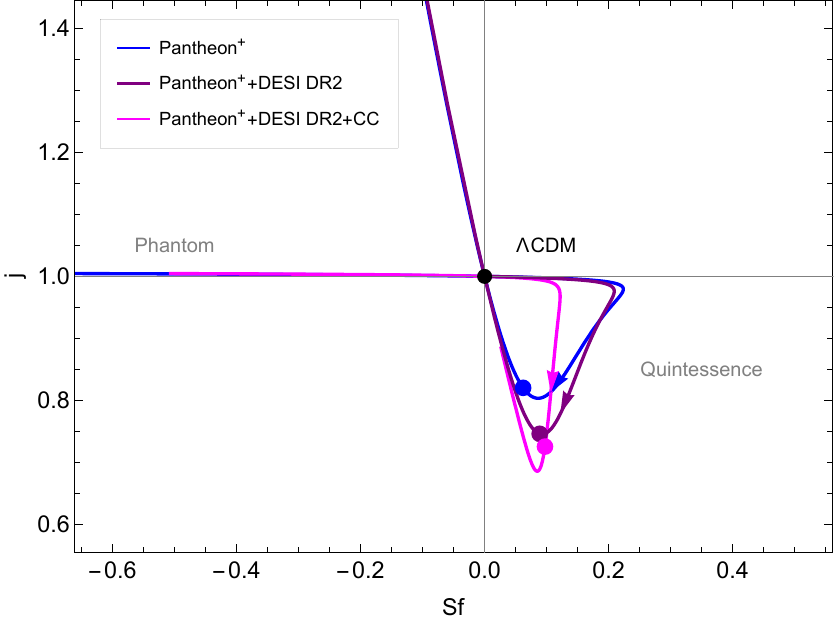}
    \caption{The evolutionary trajectories in the $\{s,j\}$ plane corresponding to the Pantheon$^{+}$, Pantheon$^{+}$+DESI DR2, and Pantheon$^{+}$+DESI DR2+CC datasets. The statefinder of the $\Lambda$CDM model appears as a fixed point, marked by a star, while the solid circles indicate the present day values of the statefinder parameters.}
    \label{fig:9}
\end{figure}

Fig.~\ref{fig:9} displays the evolutionary trajectories in the $\{s,j\}$ plane reconstructed for the power law $f(Q)$ model using the Pantheon$^{+}$, Pantheon$^{+}$+DESI DR2, and Pantheon$^{+}$+DESI DR2+CC datasets. The fixed point corresponding to the $\Lambda$CDM model, $(s=0,\,j=1)$, is marked by a star, while the solid circles indicate the present day values of the statefinder parameters for each dataset.

As shown in the figure, for all datasets, the trajectories originate in the phantom region of the $\{s, j\}$ plane and evolve toward the vicinity of the $\Lambda$CDM fixed point in the past. This transition occurs through the quintessence region, followed by a second transition toward the $\Lambda$CDM fixed point and subsequently into the phantom region in the future.

In the phantom region, before the $\Lambda$CDM fixed point, the trajectories corresponding to the Pantheon$^{+}$ and Pantheon$^{+}$+DESI DR2 datasets coincide. In the quintessence region, the trajectories evolve differently for all dataset combinations, while after the $\Lambda$CDM fixed point, they follow a similar evolution.

\section{Conclusion}
\label{sec:7}

In this work, we investigated a cosmological scenario within the framework of symmetric teleparallel $f(Q)$ gravity, considering a power law dependence on the non metricity scalar, $f(Q) = Q + 6 \gamma H_0^2 \left(\frac{Q}{Q_0}\right)^n$. The effective dark energy sector induced by non metricity was modeled through a logarithmic redshift dependent equation of state, $\omega_{\rm de}(z) = \omega_0 + \omega_1 \ln(1+z)$, allowing for a mild evolution from a cosmological constant like behavior at low redshifts.

Using recent observational datasets including Pantheon$^{+}$, DESI DR2, and Cosmic Chronometers we constrained model parameters via a Markov Chain Monte Carlo analysis. The results indicate that the Hubble parameter, $h$, and matter density, $\Omega_{m0}$, are tightly constrained and are in close agreement with $\Lambda$CDM predictions. The dark energy equation of state remains close to $\omega_0 \simeq -1$, while the parameter $\omega_1$ is generally small and may take either positive or negative values depending on the dataset combination, indicating a possible mild deviation from $\Lambda$CDM within observational uncertainties.

Cosmographic diagnostics of the deceleration ($q$), jerk ($j$), and snap ($s$) parameters reveal a smooth cosmic evolution: the Universe transitions from a past matter dominated decelerated phase to the present accelerated expansion, with higher order derivatives well behaved across all datasets. The present day values, $q_0 \simeq -0.45$, $j_0 \simeq 0.75$, and $s_0 \simeq -0.49$ (averaged over datasets), indicate that the model closely reproduces the kinematic features of $\Lambda$CDM, while allowing for controlled deviations at earlier epochs.

The reconstructed energy density parameters, $\Omega_m(z)$ and $\Omega_{\rm de}(z)$, confirm a smooth transition from matter dominance to dark energy dominance at low redshift, consistent with the observed accelerated expansion. The $\mathrm{Om}(z)$ diagnostic further supports the presence of a dynamical dark energy component, showing a mild redshift dependence below the constant $\Lambda$CDM value, particularly when combining all datasets.

The analysis of the statefinder trajectories in the $\{s,j\}$ plane for the power law $f(Q)$ model shows a consistent behavior across different datasets. The trajectories originate in the phantom region, pass through the quintessence regime, and approach the $\Lambda$CDM fixed point $(s=0,\,j=1)$ before evolving back into the phantom region in the future. While the paths coincide for Pantheon$^{+}$ and Pantheon$^{+}$+DESI DR2 datasets in the initial phantom region, the quintessence phase exhibits dataset dependent evolution, with all trajectories converging again after the $\Lambda$CDM point. This behavior highlights the model's ability to reproduce both phantom and quintessence like dynamics while remaining compatible with $\Lambda$CDM at present.

The statistical analysis using AIC$_c$ and BIC indicates that the power law $f(Q)$ model provides a competitive description of the observational data. For Pantheon$^{+}$+DESI DR2, a moderate preference is found according to AIC$_c$, while the inclusion of CC slightly weakens this preference but tightens the parameter constraints, bringing them closer to the $\Lambda$CDM limit. Although BIC penalizes the model due to its additional free parameters, the overall results demonstrate that the power law $f(Q)$ scenario remains statistically consistent and phenomenologically viable, allowing for controlled deviations from a constant dark energy equation of state while staying compatible with $\Lambda$CDM.

Overall, our analysis demonstrates that the power law $f(Q)$ model offers a consistent and phenomenologically viable description of the late time Universe. It provides a geometric alternative to dark energy, capable of reproducing the observed expansion history, and highlights the role of non metricity in shaping cosmic dynamics. These results motivate further theoretical and observational studies to explore extensions of the model and the fundamental geometric properties of gravity beyond Riemannian frameworks.

Future perspectives include the use of additional observational probes, such as redshift space distortions, and the cosmic microwave background, to test the model more rigorously and to identify potential deviations from $\Lambda$CDM, particularly at the level of perturbations. From a theoretical point of view, it is valuable to explore extensions of the $f(Q)$ model, consider alternative parameterizations of the dark energy equation of state, and analyze the evolution of cosmological perturbations and model stability to assess its robustness beyond the background evolution. These directions provide promising avenues to more clearly distinguish modified gravity scenarios from the standard cosmological model as observational data continue to improve.

\end{document}